\documentclass{PoS}
\usepackage{epsfig}
\usepackage{bm}
\usepackage{amssymb}
\usepackage{fontenc}
\usepackage{times}
\usepackage{mathptmx}
\usepackage{graphicx}

\title{Further studies of QCD with sextet quarks}

\ShortTitle{Further studies of QCD with sextet quarks}

\author{\speaker{D.~K.~Sinclair}%
         \thanks{This research was supported in part by US Department of Energy
         contracts DE-AC02-06CH11357 and \newline DE-FG02-12ER41871}
         \\
HEP Division, Argonne National Laboratory, 9700 South Cass Avenue, Argonne, 
Illinois 60439, USA\\
E-mail: \email{dks@hep.anl.gov}}

\author{J.~B.~Kogut\\
Department of Energy, Division of High Energy Physics, Washington, DC 20585,
USA\\
and\\
Department of Physics -- TQHN, University of Maryland, 82 Regents Drive, 
College Park, MD 20742, USA\\
        E-mail: \email{jbkogut@umd.edu}}

\abstract{We continue our simulations of QCD with 2 flavours of colour-sextet
quarks as a model for walking technicolor. QCD with 3 flavours of colour-sextet
quarks is also studied for comparison with the 2-flavour theory. We simulate
these theories at finite temperatures T, using lattices with a finite extent
$N_t a=1/T$ in the (Euclidean) time direction. The lattice coupling at the
chiral-symmetry-restoration transition is measured as a function of $N_t$. If
this is indeed a finite-temperature transition, the evolution of this coupling
with $N_t$ as $N_t \rightarrow \infty$ and hence the lattice spacing $a
\rightarrow 0$ should be described by asymptotic freedom. If so, the theory is
QCD-like and walking. If, however, this coupling approaches a constant
non-zero value in the large $N_t$ limit, the transition is a bulk transition
and the continuum theory is conformal. For the 2-flavour theory, the coupling
does show a significant decrease between $N_t=8$ and $N_t=12$, favouring the
walking scenario. However, preliminary results are that the change is less than
that predicted by asymptotic freedom. For the 3-flavour case, which is expected
to be conformal, there is still a significant decrease in the coupling between
$N_t=6$ and $N_t=8$, indicating that we are not yet at large enough $N_t$.}

\FullConference{31st International Symposium on Lattice Field Theory - LATTICE 
2013\\
		July 29 - August 3, 2013\\
		Mainz, Germany}

\begin{document}

\section{Introduction}

In the standard Standard Model, the Higgs field is an elementary complex scalar,
whose couplings to the $SU(2) \times U(1)$ electroweak gauge bosons 
$\gamma,W^\pm,Z$ are prescribed by gauge invariance. It has Yukawa 
couplings to the quarks and leptons, and a quartic self-coupling. Its 
quadratic self-coupling with dimensions of mass-squared is negative, so that it
develops a vacuum expectation $v$ which breaks $SU(2) \times U(1)$
spontaneously. The $W^\pm$ and $Z$ gain masses by `eating' the 3 
Goldstone bosons. $v$ gives masses to the fermions through the Yukawa 
couplings. The remaining (radial) component of the Higgs field is the so-called
Higgs particle. The recent discovery of a light Higgs candidate at the LHC
(mass~$\approx 125$~GeV) makes attempts to understand the Higgs sector of the
Standard Model timely. 

We consider the possibility that this simplest model of the Higgs sector of the
Standard Model is merely an effective field theory, and that the Higgs fields
are composite. The simplest theories of this type are Technicolor theories
\cite{Weinberg:1979bn,Susskind:1978ms}. These are QCD-like gauge theories with
massless (techni-)quarks, where the (techni-)pions play the role of the Higgs
field giving masses to the $W$s and $Z$. Technicolor theories which are merely
scaled-up QCD are not phenomenologically viable. It can be argued that Walking
Technicolor theories
\cite{Holdom:1981rm,Yamawaki:1985zg,Akiba:1985rr,Appelquist:1986an}, where the
gauge group and fermion content are such that there is a range of length/mass
scales where the running coupling evolves very slowly, might overcome these
problems \cite{Appelquist:1998xf,Hsu:1998jd,Kurachi:2006mu,Appelquist:2010xv}.
We are trying to identify gauge theories that walk. If we find such a theory,
we then need to check if it is indeed phenomenologically viable. An important
aspect of an acceptable theory is that it must describe the light ($m_H
\approx \frac{1}{2}v$) Higgs-like particle observed at the LHC. Gauge theories
with fermion content such that the theory is asymptotically free, but the 1-
and 2-loop contributions to the $\beta$-function have opposite signs, are
expected to be either conformal or walking.

Our candidate theory is (techni-)QCD with 2 massless (techni-)colour-sextet
(techni-)quarks, which could be either walking or conformal. We contrast this
theory with the 3-(techni-)flavour version, which is expected to be conformal.
QCD with 2 colour-sextet quarks has 3 Goldstone bosons -- the correct number to
give mass to the $W$s and $Z$ with none left over. In this sense it {\it is}
minimal. For other arguments as to why this theory might be of interest see
\cite{Sannino:2004qp,Dietrich:2005jn}.

Because chiral symmetry breaking and confinement occur at very different
scales in this theory, one expects (techni-)hadrons associated with both
scales. Hence light hadrons at the confinement scale could well have masses $<
f_{\pi(TC)}=v$. Thus it could potentially have light Higgs-like particles
other than the dilaton.

Other groups working on this model include DeGrand {\it et al.}
\cite{Shamir:2008pb,DeGrand:2008kx,DeGrand:2009hu,DeGrand:2010na,DeGrand:2012yq,
DeGrand:2013uha,shamir} and Fodor {\it et al.}
\cite{Fodor:2009ar,Fodor:2011tw,Fodor:2012ty,Fodor:2012ni,Fodor:2012uw,holland,
wong}.

We simulate lattice QCD with 2 sextet quarks at finite temperature, and
measure the running of the couplings at the deconfinement and chiral
transitions, as the lattice spacing is varied
\cite{Kogut:2011ty,Sinclair:2012fa}. On lattices with finite temporal extent
$N_ta$ and spatial extent $N_sa$ with $N_s >> N_t$, the temperature is
$T=1/N_ta$. If our transitions are finite temperature transitions, they will
remain at fixed temperatures as the lattice spacing $a$ is varied. If we
increase $N_t \rightarrow \infty$ at either transition, $a \rightarrow 0$. The
bare (lattice) coupling $g$ at the transition should approach zero as 
$N_t\rightarrow \infty$ in the manner described by the (perturbative)
$\beta$-function. If on the other hand, the transition is a bulk transition,
$g$ will approach a non-zero limit as $N_t \rightarrow \infty$. In this case
the field theory is conformal.  Our simulations use the Wilson (plaquette)
gauge action and the unimproved staggered fermion action. The RHMC simulation
algorithm is used to tune to a number $N_f$ of fermions which is not a
multiple of 4 ($N_f=2$).

Since the deconfinement transition occurs at a value of $\beta=6/g^2$ that
is too small to observe asymptotic freedom, for the $N_t$s we use, we 
concentrate our effort on the chiral-transition-$\beta$, $\beta_\chi$.

We simulate the $N_f=2$ theory on lattices with $N_t=4,6,8,12$ and hope to
extend this to larger $N_t$. Preliminary results indicate that
$\beta_\chi(N_t=12)$ is significantly larger than $\beta_\chi(N_t=8)$, but by
less than what the 2-loop $\beta$-function would predict.

We simulate the $N_f=3$ theory on lattices with $N_t=4,6,8$ and will extend
this to $N_t=12$ \cite{Kogut:2011bd,Sinclair:2012fa}. Preliminary results
indicate that $\beta_\chi(N_t=8)$ is significantly greater than
$\beta_\chi(N_t=6)$ which would indicate that we are not yet at large enough
$N_t$, since we expect $\beta_\chi$ to approach a non-zero constant at large
$N_t$. Even if this theory is QCD-like, asymptotic freedom would predict that
the change in $\beta_\chi$ between $N_t=6$ and $N_t=8$ should be too small to
be measured in our current simulations. Again we simulate using the RHMC
algorithm since $N_f=3$ is not a multiple of $4$.

\section{Lattice QCD with 2 flavours of colour-sextet quarks}

\subsection{Simulations with $N_t=8$}

Our $N_t=8$ simulations were performed primarily on a $16^3 \times 8$ lattice.
Simulations performed on a $24^3 \times 8$ lattice at $\beta=6.7$ and
$\beta=6.9$, both with $m=0.0025$ (the smallest mass we use, indicate that
finite lattice size errors are small on lattices of spatial extent $N_s=16$.
We have no new $16^3 \times 8$ results since Lattice 2012
\cite{Sinclair:2012fa}.

Our $16^3 \times 8$ simulations were performed at $m=0.02$, $m=0.01$,
$m=0.005$, and $m=0.0025$ to allow performance of chiral extrapolations. In
the neighbourhood of the chiral transition, $6.6 \le \beta \le 6.8$, we
performed runs of 50,000 length-1 trajectories at $m=0.02$, $m=0.01$, and
$m=0.005$ for each $\beta$ and $m$ for $\beta$s spaced at $0.02$ intervals. At
$m=0.0025$ we have performed runs of 100,000 trajectories for each $\beta$ in
this range.

Since it is difficult if not impossible to extrapolate either the unsubtracted
or subtracted chiral condensates to zero quark mass with sufficient reliability
to determine the position of the chiral-symmetry-restoration phase transition
accurately, we estimate the value of $\beta_\chi$ from the peaks in the
(disconnected) chiral susceptibility:
\begin{equation}
\chi_{\bar{\psi}\psi} = \frac{V}{T}\left[\langle (\bar{\psi}\psi)^2 \rangle
                                   -\langle \bar{\psi}\psi \rangle^2\right]
\end{equation}
extrapolated to $m=0$. Since the positions of the peaks in these
susceptibilities show little mass dependence, we take the position of the peak
for $m=0.0025$ as our estimate of $\beta_\chi$. This gives
$\beta_\chi=6.69(1)$. Ferrenberg-Swendsen reweighting was used to interpolate
between $\beta$ values to determine this peak. We refer to our Lattice 2012
talk \cite{Sinclair:2012fa} for graphs. For $m=0.0025$ and $\beta=6.7$, and for
$\beta=6.9$, these susceptibilities are consistent with those measured on a
$24^3 \times 8$ lattice.

\subsection{Simulations with $N_t=12$}

\begin{figure}[htb]
\parbox{2.9in}{
\epsfxsize=2.9in
\centerline{\epsffile{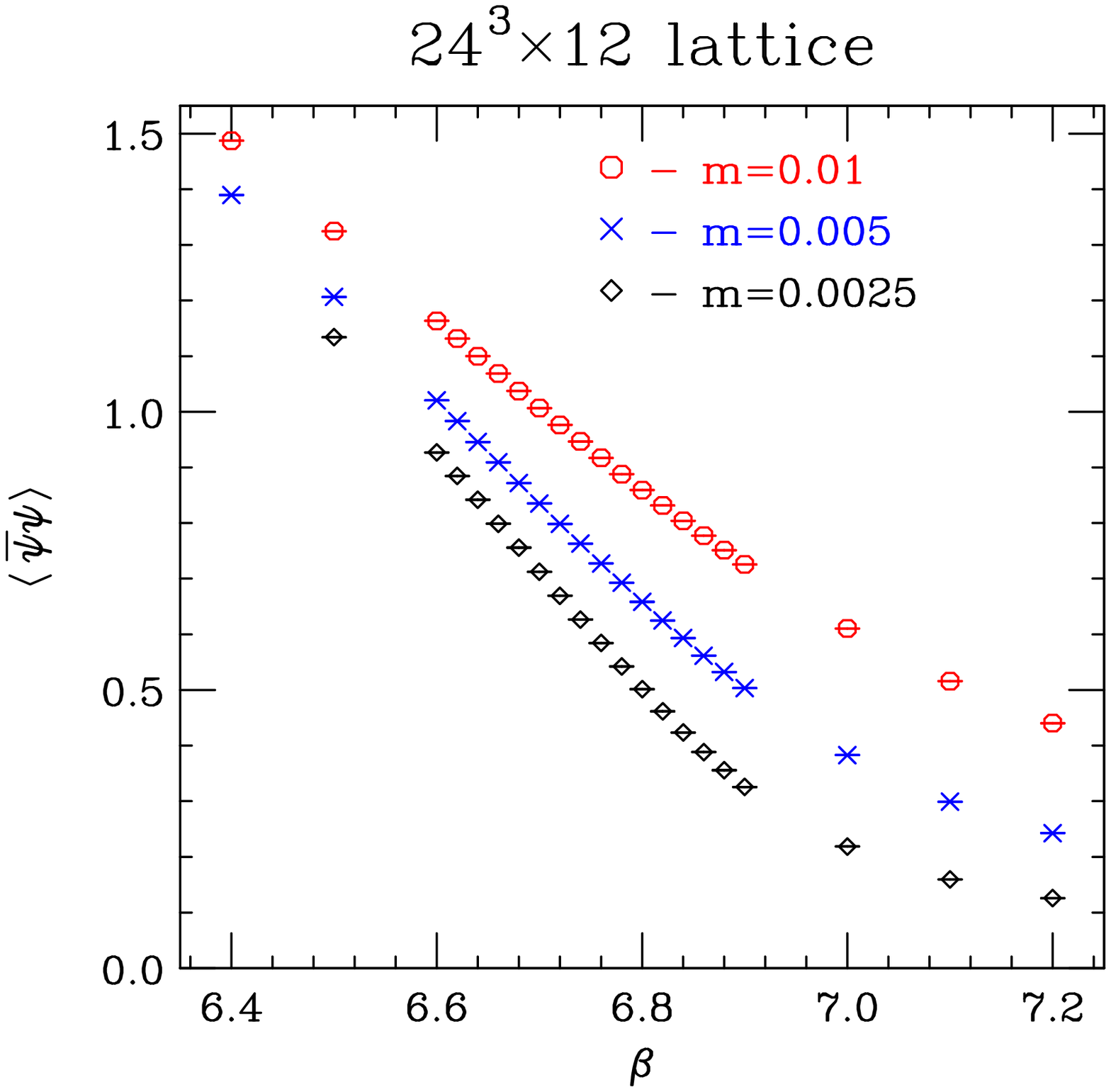}}
\caption{Chiral condensates on a $24^3 \times 12$ lattice.}
\label{fig:pbp12}
}
\parbox{0.2in}{}
\parbox{2.9in}{
\epsfxsize=2.9in    
\centerline{\epsffile{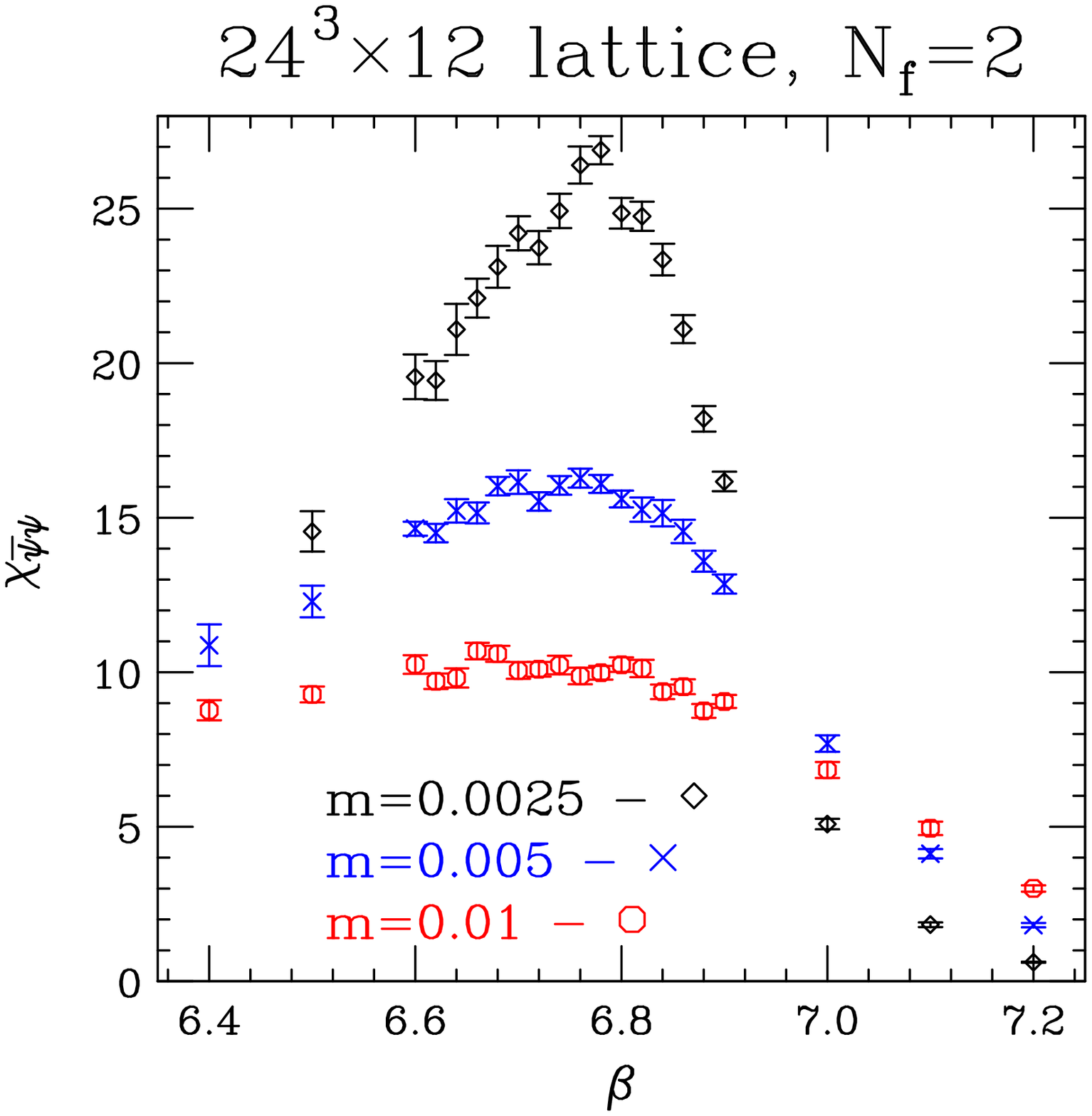}}
\caption{Chiral susceptibilities on a $24^3 \times 12$ lattice.}
\label{fig:chi12}
}
\end{figure}

We are now extending our simulations at $N_t=12$ on $24^3 \times 12$
lattices, with quark masses $m=0.01$, $m=0.005$, and $m=0.0025$. In the
neighbourhood of the chiral transition $6.6 \le \beta \le 6.9$ we perform
simulations at $\beta$s spaced by $\delta\beta=0.02$. So far we have simulated
20,000 -- 100,000 length-1 trajectories at each $(\beta,m)$ in this range.
Figure~\ref{fig:pbp12} shows the unsubtracted chiral condensates
$\langle\bar{\psi}\psi\rangle$ measured in these simulations. Note that,
although these suggest that the condensate will vanish in the chiral limit for
$\beta$ sufficiently large, any attempt to extrapolate to $m=0$ to estimate
$\beta_\chi$ would be plagued with systematic uncertainties. Using subtracted
condensates improves the situation, but not sufficiently to extract
$\beta_\chi$ with the needed precision.

To determine the position $\beta_\chi$ of the chiral transition with
sufficient accuracy, we examine the peak in the (disconnected) chiral
susceptibility as a function of mass. These susceptibilities are shown in
figure~\ref{fig:chi12}. Both the $m=0.005$ and $m=0.0025$ susceptibilities
show peaks. While the `data' is consistent with there being little mass
dependence of the position of the peaks, it is not yet compelling. More
statistics is needed. Our best estimate of $\beta_\chi$ from the current data
is $\beta_\chi=6.78(2)$. This implies that:
\begin{equation}
\beta_\chi(N_t=12)-\beta_\chi(N_t=8) = 0.09(2)
\end{equation}
compared with the 2-loop perturbative prediction
\begin{equation}
\beta_\chi(N_t=12)-\beta_\chi(N_t=8) \approx 0.12
\end{equation}

\subsection{Simulations on $24^3 \times N_t$ lattices}

In addition to our fixed $N_t$ simulations, we simulate this 2-flavour theory
on $24^3 \times N_t$ lattices with $N_t \le 24$ at fixed $\beta$, to search
for evidence of a transition back to the chirally broken phase, as $N_t$ is
increased. We choose $\beta=6.9$. This beta was chosen since, if the evolution
of $\beta_\chi$ is governed by the 2-loop $\beta$-function, $\beta_\chi
\approx 6.9$ for $N_t=18$. Hence we should see evidence for the chiral
transition as $N_t$ is increased.

\begin{figure}[htb]
\parbox{2.9in}{
\epsfxsize=2.9in
\centerline{\epsffile{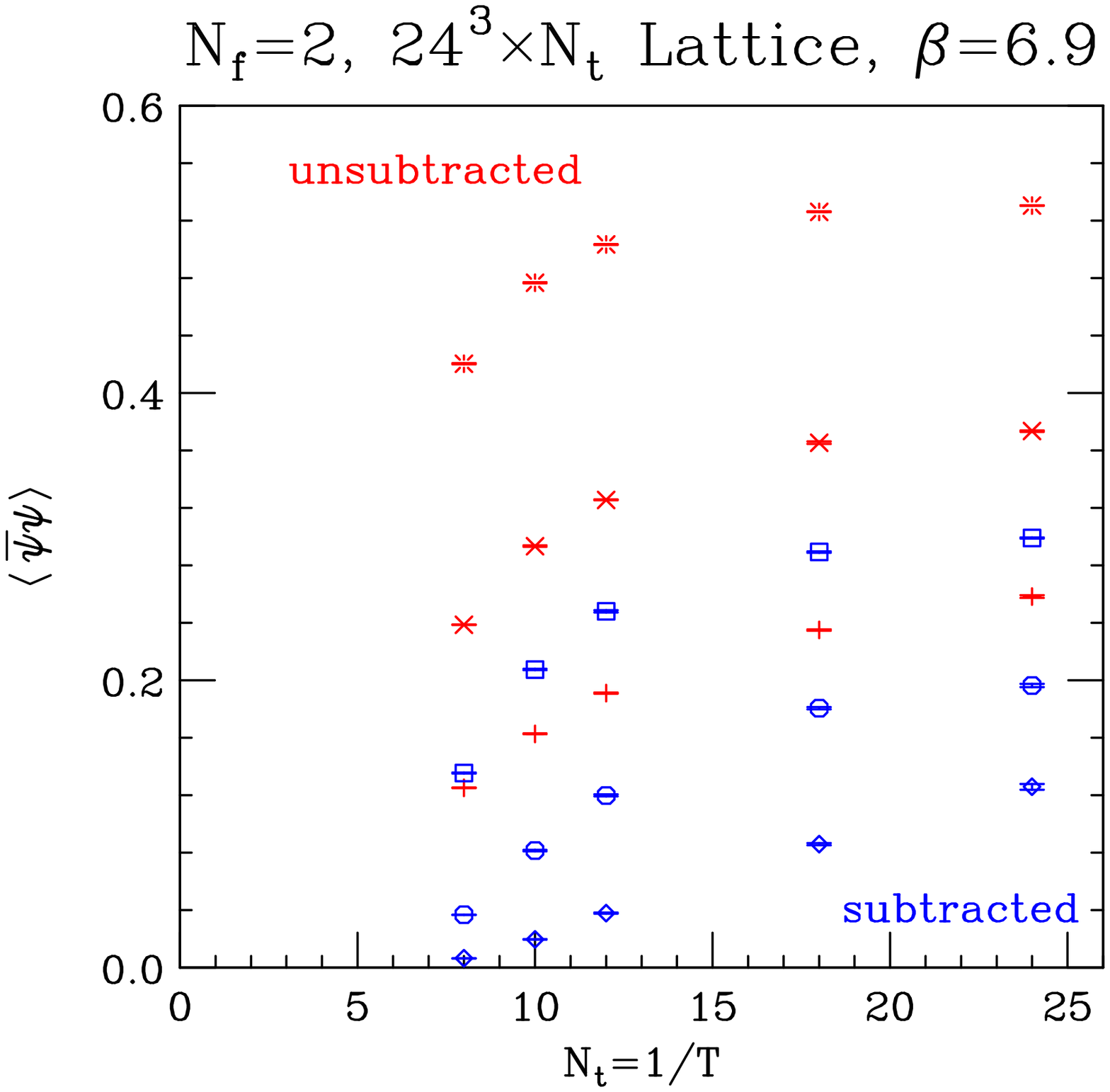}}                        
\caption{Chiral condensates on $24^3 \times N_t$ lattices at $\beta=6.9$. From
top to bottom, the masses are $m=0.005$, $m=0.0025$ and $m=0.00125$.}
\label{fig:pbp24Nt}                
}
\parbox{0.2in}{}
\parbox{2.9in}{
\epsfxsize=2.9in  
\centerline{\epsffile{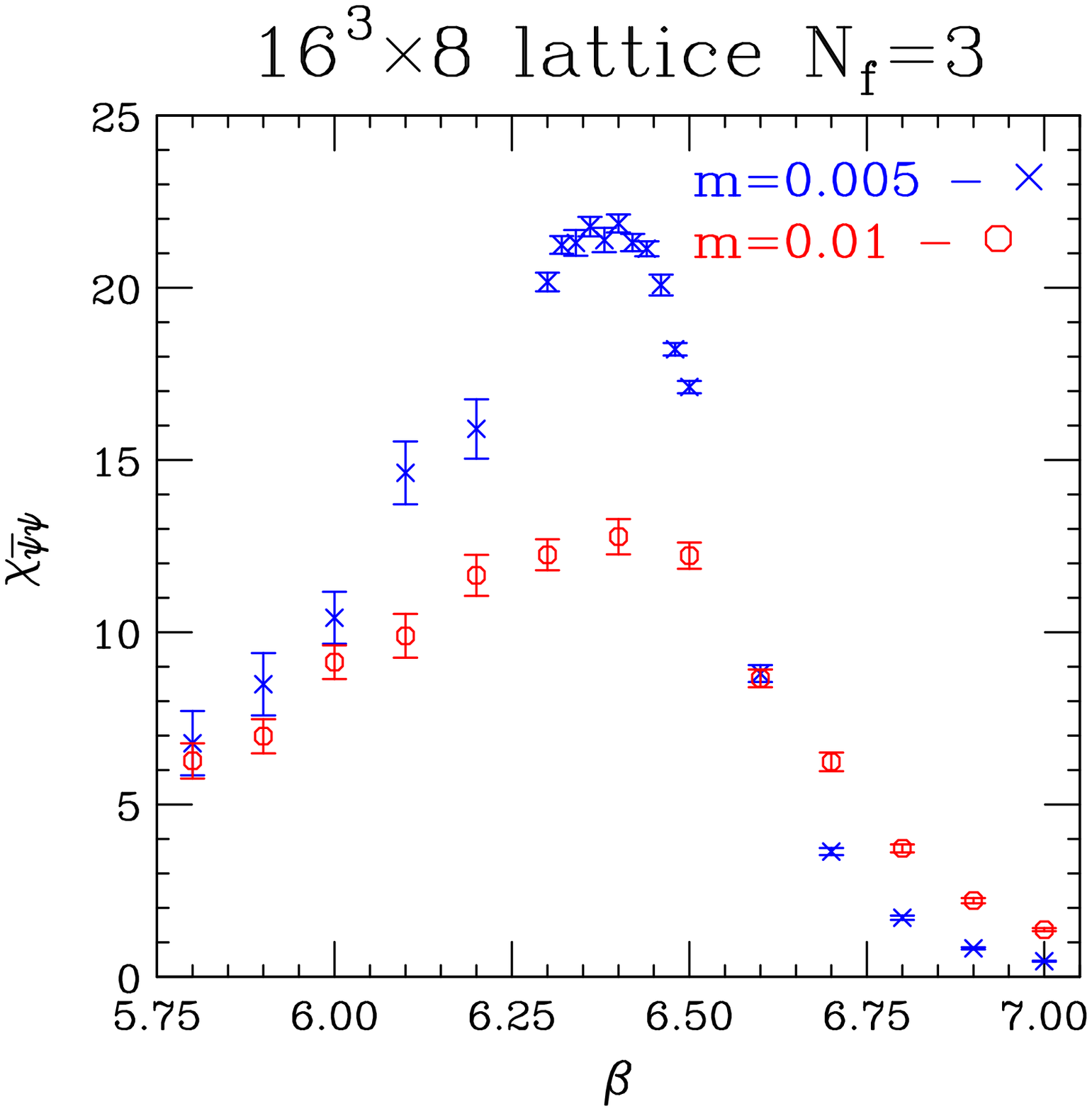}}
\caption{Chiral susceptibilities for $N_f=3$ on a $16^3 \times 8$ 
lattice.}
\label{fig:chi3_8}
}
\end{figure}

At present, we run on $24^3 \times 8$, $24^3 \times 10$, $24^3 \times 12$,
$24^3 \times 18$, $24^3 \times 20$, $24^3 \times 22$ and $24^4$ lattices at
$m=0.005$, $m=0.0025$ and $m=0.00125$. We plan runs on $24^3 \times 14$ and
$24^3 \times 16$ lattices. Figure~\ref{fig:pbp24Nt} shows the unsubtracted and
subtracted chiral condensates as functions of $N_t$ for these runs. We follow
Fodor {\it et al.}, defining a subtracted chiral condensate by:
\begin{equation}
\langle{\bar{\psi}\psi}\rangle_{sub} = \langle{\bar{\psi}\psi}\rangle 
-\left(m_V\frac{\partial}{\partial m_V}\langle{\bar{\psi}\psi}\rangle\right)
                                                                  _{m_V=m}
\end{equation}
where $m_V$ is the valence quark mass \cite{Fodor:2011tw}. 
We should try other schemes, since
this data is inconclusive. We have also looked at the chiral susceptibilities,
but with our present statistics we have not seen a peak.

\section{Lattice QCD with 3 flavours of colour-sextet quarks}

We simulate lattice QCD with 3 light quark flavours on $16^3 \times 8$
lattices. Our simulations are performed at $m=0.01$ and $m=0.005$. For 
$6.3 \le \beta \le 6.5$, which is close to the chiral transition, and $m=0.005$,
we have performed runs of 100,000 trajectories at intervals of $0.02$
in $\beta$ at each $\beta$ in this range. Figure~\ref{fig:chi3_8} shows the
chiral susceptibilities for these runs. From this we estimate that
$\beta_\chi=6.38(2)$, somewhat larger than that for $N_t=6$.

\section{Discussions and Conclusions}

\begin{table}[htb]
\centerline{
\begin{tabular}{|c|c|c|}
\hline
$N_t$          & $\beta_d$       & $\beta_\chi$             \\
\hline
4              &$\;$5.40(1)$\;$  &$\;$6.3(1)$\;$            \\
6              &$\;$5.54(1)$\;$  &$\;$6.60(2)$\;$           \\
8              &$\;$5.65(1)$\;$  &$\;$6.69(1)$\;$           \\
12             &$\;$5.8(1) $\;$  &$\;$6.78(2)$\;$           \\
\hline
\end{tabular}
}
\caption{$N_f=2$ deconfinement and chiral transitions for $N_t=4,6,8,12$.}
\label{tab:trans}
\end{table}

We are simulating lattice QCD with 2 light colour-sextet quarks at finite
temperature, to test whether it has an infrared fixed-point and is thus
conformal, or if it is QCD-like, but with a slowly-evolving running coupling
constant, i.e. if it `walks' and is thus a Walking-Technicolor candidate.
We have extended our simulations at $N_t=12$. As seen in table~\ref{tab:trans}
above, the chiral phase transition moves to larger $\beta$ as $N_t$ is
increased. However, $\beta_\chi(N_t=12)-\beta_\chi(N_t=8) = 0.09(2)$
compared with $\approx 0.12$ predicted using the 2-loop perturbative 
$\beta$-function. More statistics is needed to ratify this result. We will
also test if this is a finite size effect. It is possible that the 3-loop term
in the $\beta$-function is large in this lattice regularization. Larger $N_t$s
might be needed to clarify this issue. Note that the chiral susceptibility is
very sensitive to long `time'-constant modes describing the system's
evolution. These often have small amplitudes, so that they are not evident in
other observables.

Our simulations of the 3-flavour theory at $N_t=8$ indicate that the increase
in $\beta$, $\beta_\chi(N_t=8)-\beta_\chi(N_t=6)$ is appreciably larger than
the $\approx 0.0025$ predicted by 2-loop perturbation theory. Nor can we see
any evidence that $\beta_\chi(N_t)$ is approaching a non-zero constant as 
$N_t\rightarrow \infty$ as expected, since the this theory is expected to be
conformal. We will therefore need to perform simulations at larger $N_t$.
$N_t=12$ simulations are planned for 2014. We will also extend our $N_t=6$
simulations to obtain a better estimate of the position of its chiral phase
transition.

To study the spectrum of the 2-flavour theory at zero temperature, we need to
restrict ourselves to the region $\beta < \beta_d$, the $\beta$ at the
deconfinement transition. For this $\beta$ to also lie in the weak-coupling
domain, so that we can make contact with the continuum will require rather
large lattices. Since our experience with the chiral transition suggests that
the crossover from strong- to weak-coupling occurs somewhere in the regime
$\beta=$~6.3--6.6 while $\beta_d(N_t=12) \approx 5.75$, we estimate that we
will require lattices whose smallest dimension $N_s=48$ or larger. For zero
temperature physics, $f_{\pi(TC)}$ is the relevant scale. We know that
$f_{\pi(TC)} = v \approx 246$~GeV, so 
$m_{\rm Higgs} \approx \frac{1}{2}f_{\pi(TC)}$. This puts a severe constraint
on any model of Higgs dynamics, something that we will test in our model.

We also plan to simulate other candidate Walking Technicolor models. For
example, we are planning to study is $SU(2)$ gauge theory with 3 adjoint
Majorana/Weyl quarks, starting in 2014. A good summary of the current state of
lattice simulations of various candidate theories is contained in the review
talk by Julius Kuti at Lattice 2013 \cite{kuti}.

\section*{Acknowledgements}
These simulations are performed on Hopper, Carver and Edison at NERSC, Kraken 
at NICS, and Fusion and Blues at LCRC, Argonne. We thank Julius Kuti for
helpful insights, in particular his suggestion that we perform simulations at
a fixed $\beta$ and $N_s$, for a range of $N_t$ values.

\end{document}